\documentclass[traditabstract]{aa}
\usepackage{aecompl}

\usepackage{array}
\usepackage{amsmath}
\usepackage{graphicx}
\usepackage{amssymb}
\usepackage{widetext}
\usepackage{flushend}
\usepackage{cuted}

\usepackage[round]{natbib}

\usepackage{color}
\usepackage[colorlinks]{hyperref}
\usepackage{calc}
\usepackage{feynmf}
\usepackage{subfigure}

\providecommand{\aap}{Astron.Astrophys.}
\providecommand{\apjs}{Astrophys.J.Suppl.}
\providecommand{\apjl}{Astrophys.J.}
\providecommand{\physrep}{Phys. Rep.}
\providecommand{\mnras}{Mon. Not. Roy. Astron. Soc.}
\providecommand{\npa}{Nucl. Phys. A}
\providecommand{\plb}{Phys.~Lett. B}
\providecommand{\epja}{EPJ A}

\makeatletter

\newcommand*\@dblLabelI {}
\newcommand*\@dblLabelII {}
\newcommand*\@dblequationAux {}

\def\@dblequationAux #1,#2,%
    {\def\@dblLabelI{\label{#1}}\def\@dblLabelII{\label{#2}}}

\newcommand*{\doubleequation}[3][]{%
    \par\vskip\abovedisplayskip\noindent
    \if\relax\detokenize{#1}\relax
       \let\@dblLabelI\@empty
       \let\@dblLabelII\@empty
    \else 
       \@dblequationAux #1,%
    \fi
    \makebox[0.5\linewidth-3.5em]{%
     \hspace{\stretch2}%
     \makebox[0pt]{$\displaystyle #2$}%
     \hspace{\stretch1}%
    }%
    \makebox[0.5\linewidth+0.5em]{%
     \hspace{\stretch1}%
     \makebox[0pt]{$\displaystyle #3$}%
     \hspace{\stretch2}%
    }%
    \makebox[3em][r]{(%
  \refstepcounter{equation}\theequation\@dblLabelI,
  \refstepcounter{equation}\theequation\@dblLabelII)}%
  \par\vskip\belowdisplayskip
}

\begin{document}

\bibliographystyle{aa}

\title{Role of medium modifications for neutrino-pair processes from nucleon-nucleon bremsstrahlung}
\subtitle{Impact on the protoneutron star deleptonization}

\author{Tobias Fischer\inst{1}\thanks{{\em email:} fischer@ift.uni.wroc.pl}}
\institute{Institute for Theoretical Physics, University of Wroc{\l}aw, plac Maksa Borna 9, 50-204 Wroc{\l}aw, Poland}

\date{\today}

\abstract
{In this article the neutrino-pair production from nucleon-nucleon ($NN$) bremsstrahlung is explored via medium-modifications of the strong interactions at the level of the one-pion exchange approximation. It governs the bulk part of the $NN$ interaction at low densities relevant for the neutrino physics in core-collapse supernova studies. The resulting medium modified one-pion exchange rate for the neutrino-pair processes is implemented in simulations of core collapse supernovae in order to study the impact on the neutrino signal emitted from the deleptonization of the nascent proto-neutron star. Consequences for the nucleosynthesis of heavy elements of the material ejected from the PNS surface are discussed.}


\maketitle

\section{Introduction}
\label{intro}
The bulk part of trapped neutrinos of a newly born neutron star -- known as proto-neutron star (PNS) -- is emitted during its early cooling history. PNSs form when the stellar core of an initially imploding massive star bounces back at supersaturation density, due to the repulsive short range nuclear interaction, with the formation of a strong hydrodynamic shock wave \citep[for recent review about supernova theory, cf.,][]{Janka:2007}. While the supernova problem is associated with the transport of energy from the PNS interior leading to the ejection of the stellar mantle --  a detailed summary about supernova explosion mechanisms can be found in \citet{Janka:2012} -- the nascent PNS deleptonizes via the emission of neutrinos of all flavors on a timescale on the order of 10--30~s once the supernova explosion has been launched. This has been explored in \citet{Fischer:2009af} and \citet{Huedepohl:2010} within consistent simulations based on neutrino radiation hydrodynamics with three-flavor Boltzmann neutrino transport. These studies confirmed that the PNS settles into a quasi-stationary state as mass accretion ceases, which approves the description within the hydrostatic treatment of \citet{Pons:1998mm} based on the diffusion approximation for neutrino transport.

In \cite{Wu:2015} for the first time the ''complete'' supernova neutrino signal has been analyzed based on the simulation results of \citet{Fischer:2009af}. It demonstrated that millisecond events can be resolved for the next Galactic supernova explosion, for currently operating and for future generation of neutrino detectors. Therefore, it is of paramount interest to predict neutrino luminosities and spectra from such cosmic events. Particular focus is thereby on the deleptonization of the nascent PNS during which most neutrinos will be detected. From a theory point of view it is important to develop models that are based on accurate neutrino transport considering the relevant weak processes. Following \citet{Reddy:1998} the impact of the consistent treatment of nuclear equation of state (EoS) and weak processes has been explored at the mean-field level in \citet{MartinezPinedo:2012}, \citet{Roberts:2012} and \citet{Horowitz:2012}.

\setlength{\unitlength}{1.1mm}
\begin{figure}[b!]
\begin{center}
\bigskip
\begin{fmffile}{feynman}
\begin{fmfchar*}(35,25)
\fmfleftn{l}{2}
\fmfrightn{r}{4}
\fmfforce{(0.0w,0.2h)}{l1}
\fmfforce{(0.0w,0.8h)}{l2}
\fmfforce{(1.2w,0.2h)}{r1}
\fmfforce{(1.2w,0.8h)}{r2}\fmfforce{(1.2w,1.1h)}{r3}\fmfforce{(1.2w,1.3h)}{r4}
\fmfforce{(0.9w,1.1h)}{o2}\fmfforce{(0.9w,0.8h)}{o1}
\fmfforce{(0.5w,0.8h)}{ul}
\fmfforce{(0.5w,0.8h)}{ur}
\fmfforce{(0.5w,0.84h)}{uo}
\fmfforce{(0.5w,0.2h)}{dl}
\fmfforce{(0.5w,0.2h)}{dr}
\fmfforce{(0.5w,0.2h)}{do}
\fmf{fermion}{l2,ul}
\fmf{fermion,label=$N$,label.side=left}{ur,ddl}
\fmf{fermion,label=$N$,label.side=right}{ddr,r2}
\fmf{fermion}{l1,dl}
\fmf{fermion}{dr,r1}
\fmf{dashes,label=$\pi$}{do,uo}
\fmf{fermion}{r3,o2}
\fmf{fermion,label=$\nu$,label.side=left}{o2,r4}
\fmf{boson,label=$Z^0$}{o2,o1}
\fmf{fermion}{ur,ddl,r2}
\fmflabel{$\bar{\nu}$}{r3}
\fmflabel{$N$}{r1}\fmflabel{$N$}{l1}\fmflabel{$N$}{l2}
\end{fmfchar*}
\end{fmffile}
\\
(Time direction from left to right.)
\end{center}
\caption{Neutrino pair-emission from $NN$--bremsstrahlung within the FOPE approximation.}
\label{fig:feynman}
\end{figure}
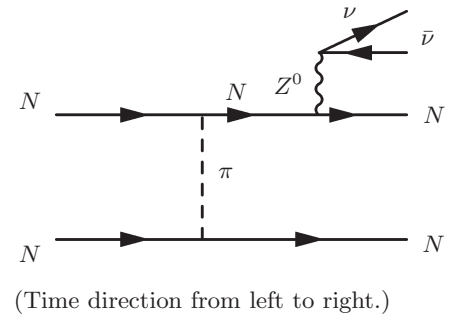

Physics beyond the mean-field approach involves nuclear correlations, such as nucleon-nucleon ($NN$) scattering and pairing. However, $NN$ interactions under supernova conditions -- temperatures on the order to tens of MeV and densities slightly above normal nuclear matter density ($\rho_0$) as well as large neutron excess -- are highly unknown. This is then reflected in large uncertainties of associated weak processes, e.g., $NN$--bremsstrahlung and the corresponding neutrino-antineutrino pair emission (or absorption) as illustrated in Fig.~\eqref{fig:feynman}. The different scales for strong and weak interactions involved allow their separation; thereby the latter refers to the emission (and absorption) of $(\nu,\bar\nu)$ via the coupling to the neutral $Z^0$-boson. Therefore, \citet{Friman:1979} derived expressions for the weak rates modeling $NN$ interactions by the one-pion exchange (OPE) approximation, including spin-orbit coupling. Moreover, \citet{Hannestad:1997gc} provided expressions in the long-wavelength limit to the astrophysics community \citep[see also][and references therein]{Sigl:1995ac,Janka:1996} which take into account spin fluctuations and their saturation towards high density.

It has long been realized in a series of studies that the free (i.e. vacuum) OPE -- henceforth denoted as FOPE -- underestimates the cooling of neutron stars. Based on FOPE important medium modifications at $\rho\gtrsim \rho_0/2$ are neglected \citep[this has been discussed in details in][exploring also the role of medium effects for the charged-current absorption reactions, i.e. the modified Urca processes]{Voskresensky:1987,Blaschke:1995,Voskresensky:1997, Yakovlev:2001,Reddy:2001,Blaschke:2004,Yakovlev:2005,Grigorian:2005,Reddy:2006,Blaschke:2012,Blaschke:2013}.

Up to $\rho_0$ the framework of chiral effective field theory is the ab-initio approach to the nuclear many body problem of dilute neutron matter. Therefore, \citet{Hebeler:2010a} studied the role of three-body forces including $2\pi$-exchange terms \citep[for recent works, see also][]{Tews:2013,Krueger:2013}. Based on this approach, recently \citet{Bartl:2014} quantified the suppression of the neutrino-pair emission/absorption rate from $NN$--bremsstrahlung, at conditions relevant for supernovae. The authors found a significant reduction of the neutrino-pair process compared to the FOPE approximation, at already low density on the order of $\rho\simeq 10^{11}-10^{12}$~g~cm$^{-3}$.

A more general approach to study medium modifications for weak processes that involve $NN$ interactions is the Fermi-liquid theory of \citet{Migdal:1978} and \citet{Migdal:1990} \citep[see also][and references therein]{Voskresenskaya:2001}. It assumes that the nucleons are only slightly excited above their Fermi sea such that all the processes occur in a narrow vicinity of the Fermi surface and that the vertices of the processes are dressed by $NN$ correlations. The subsequent medium modification of the OPE includes the ''exact'' summation of all particle-hole loops and a contribution of the residual $s$-wave $\pi NN$ interaction and $\pi\pi$-scattering, which effectively involves $\rho$-meson exchange ($t$-channel) and the $\sigma$-meson as correlated $\pi\pi$ ($s$-wave contribution). Moreover, due to nucleon--hole (and also $\triangle$--hole) excitations, the subsequent $p$-wave pion polarization in nuclear matter leads to the {\em softening} of the in-medium pion dispersion relation. As explored in \citet{Wambach:1995} and \citet{Rapp:1996}, it is relevant for the understanding of nuclear saturation \citep[for more details, cf.,][and references therein]{Rapp:1997}.

In this paper medium modifications to the FOPE approach are studied, which alter FOPE already at low density ($\rho\ll \rho_0$) where they start to dress the local $\pi NN$-vertex (henceforth denoted as MVOPE -- medium-modified-vertex + FOPE). Being mostly interested in the low-density limit and to easier compare with results based on the chiral perturbation approach, where pion softening is not included, medium effects for the $\pi NN$-vertices are incorporated but the FOPE $\pi$-propagator is exploited.

Following \citet{Voskresensky:1987} and \citet{Voskresenskaya:2001} the current article provides a MVOPE density-dependent parametrization, i.e. for the medium modification of the $\pi NN$-vertex, for applications in simulations of core-collapse supernovae. It represents the largest modification of the neutrino-pair emission rate from $NN$--bremsstrahlung at densities relevant for supernova studies,  suppressing neutrino-pair emission/absorption  compared to the unmodified FOPE approach.

This medium dependence is then implemented into the neutrino transport module of the core-collapse supernova model AGILE-BOLTZTRAN of \citet{Liebendoerfer:2004} in order to study the impact on the neutrino signal and the evolution during the PNS deleptonization, being able to quantify the general role of $NN$--bremsstrahlung. The reduction of the neutrino-pair emission/absorption rate obtained towards increasing density reduces the neutrino opacity. This is important for the heavy-lepton flavor neutrino species for which $NN$--bremsstrahlung is the leading production and absorption process. Unlike for $\nu_e$ which are always dominated by charged-current absorption on neutrons, also $\bar\nu_e$ are significantly affected from the high-density suppression of the $NN$--bremsstrahlung.

The subsequent impact on the neutrino luminosities and average energies has important consequences for the nucleosynthesis relevant conditions, e.g., the electron fraction $Y_e$ i.e. the proton-to-baryon ratio, of the neutrino-driven wind. The latter is a low-mass outflow ejected from the PNS surface via continuous neutrino heating, during the PNS deleptonization over a timescale of about 10--30~seconds. It has long been studied as nucleosynthesis site for the production of heavy elements \citep[cf.][and references therein]{Woosley:1994ux,Otsuki:1999kb,Wanajo:2006mq,Wanajo:2006ec}.

Note that for $\rho\gtrsim\rho_0$ also the $\pi$-propagator is modified by the nuclear medium, which gives rise to different interaction channels \citep[cf. the right panel of Fig.~(1) in][]{Voskresenskaya:2001} with neutrino-pair emission from intermediate nucleon-loops. Such additional medium modifications may become relevant during the late-time cooling at the transition from neutrino diffusion to freely streaming, when the PNS core temperature decreases below $T\sim 1$~MeV such that neutrinos are escaping from densities also in excess of $\rho_0$.

The manuscript is organized as follows. In sec.~\ref{SNmodel} the supernova model AGILE-BOILTZTRAN is briefly introduced together with the density-dependent parametrization of medium modifications for the neutrino-pair processes from $NN$--bremsstrahlung. In sec.~\ref{PNSevol} simulation results of the PNS deleptonization are analyzed and in sec.~\ref{ndw} the impact on the nucleosynthesis relevant conditions is discussed. The manuscript closes with the summary in sec.~\ref{summary}.

\begin{table*}[htp!]
\centering
\caption{List of weak processes considered and corresponding references.}
\begin{tabular}{ccc}
\hline
\hline
& Weak process & References \\
\hline
1 & $e^- + p \rightleftarrows n + \nu_e$ & \citet{Reddy:1998,Horowitz:2001xf} \\ 
2 & $e^+ + n \rightleftarrows p + \bar\nu_e$ & \citet{Reddy:1998,Horowitz:2001xf} \\
3 & $n \rightleftarrows p + e^- + \bar\nu_e$ & \citet{Fischer:2016b} \\
4 & $e^- + (A,Z) \rightleftarrows (A,Z-1) + \nu_e$ & \citet{Juodagalvis:2010} \\
5 & $\nu + N \rightleftarrows  N + \nu'$ & \citet{Bruenn:1985en,Mezzacappa:1993gm,Horowitz:2001xf} \\
6 & $\nu + (A,Z) \rightleftarrows (A,Z) + \nu'$ & \citet{Bruenn:1985en,Mezzacappa:1993gm} \\
7 & $\nu + e^\pm \rightleftarrows e^\pm + \nu'$ & \citet{Bruenn:1985en,Mezzacappa:1993gx} \\
8 & $e^- + e^+ \rightleftarrows  \nu + \bar{\nu}$ & \citet{Bruenn:1985en} \\
9 & $N + N \rightleftarrows  N + N + \nu + \bar{\nu}$ & \citet{Hannestad:1997gc} \\
10 & $\nu_e + \bar\nu_e \rightleftarrows  \nu_{\mu/\tau} + \bar\nu_{\mu/\tau}$ & \citet{Buras:2002wt,Fischer:2009} \\
11 & $(A,Z)^* \rightleftarrows (A,Z) + \nu + \bar\nu$ & \citet{Fuller:1991,Fischer:2013} \\
\hline
\end{tabular}
\\
Note: unless stated otherwise, $\nu=\{\nu_e,\bar{\nu}_e,\nu_{\mu/\tau},\bar{\nu}_{\mu/\tau}\}$ and $N=\{n,p\}.$
\label{tab:nu-reactions}
\end{table*}
%

\section{Supernova input physics}
\label{SNmodel}
The core-collapse supernova model, AGILE-BOLTZTRAN, is based on spherically symmetric and general relativistic neutrino radiation hydrodynamics with angle- and energy-dependent three flavor Boltzmann neutrino transport~\citep[for details, cf.][]{Liebendoerfer:2001a,Liebendoerfer:2001b,Liebendoerfer:2002}. The implicit method for solving the hydrodynamics equations and the Boltzmann equation on an adaptive Lagrangian mesh has been compared with other methods, e.g., in \citet{Liebendoerfer:2004} with the multi-group flux limited diffusion approximation and in \citet{Liebendoerfer:2005a} with the variable Eddington factor technique.

The nuclear EoS employed here from \citet{Hempel:2009mc} is based on the relativistic mean-field framework for homogeneous nuclear matter with the parametrization DD2 from {\citet{Typel:2009sy}, henceforth denoted as HS(DD2). Nuclei are treated within the modified nuclear statistical equilibrium approach for several 1000 nuclear species based on tabulated and partly calculated masses. It is part of the comprehensive CompOSE EoS catalogue of \citet{Typel:2013rza}. In addition lepton and photon contributions are added following \citet{Timmes:1999}.

\subsection{Treatment of weak processes}
The set of weak reactions included in the supernova simulations can be found in Table~\ref{tab:nu-reactions}, together with the corresponding references. For the weak processes with free nucleons, both charged-current absorption and neutral current scattering, the elastic approximation is employed. Medium modifications of the charged current absorption processes are taken into account at the mean-field level based in the nuclear EoS HS(DD2). Following \citet{Reddy:1998}, these enter the elastic expressions via modified $Q$ values, $Q=Q_0+\triangle U$, where $\triangle U = U_n - U_p$ with nucleon single-particle self energies $U_N$. They also modify the nucleon chemical potentials, $\mu_N=\varphi_N-U_N$, with free Fermi-gas chemical potentials $\varphi_N$, in order to reproduce the low-density limit towards vanishing degeneracy \citep[cf.][for details regarding the treatment of effective interactions within relativistic mean-field approach]{Hempel:2015b}. Note that $U_N$ have a strong density dependence, which is related to the nuclear symmetry energy. Even though it was possible to constrain the symmetry energy value and its slope to some extend in \citet{Lattimer:2013}, under supernova conditions the symmetry energy is presently rather poorly constrained as discussed in \citet{Fischer:2014}. 

Furthermore, inelastic contributions and weak magnetism corrections are included following \citet{Horowitz:2001xf}, for the charged current absorption and neutral current scattering processes. While inelastic contributions are known to reduce the $\nu$ rates as much as $\bar\nu$ rates towards high density, weak magnetism corrections generally increase differences between neutrinos and antineutrinos. The latter originates from the parity violation of the nucleon weak magnetic moment coupling to its axial current. Both latter effects have been commonly included in core-collapse supernova simulations. Note that contributions due to strange quark contents are not taken into account here.

In addition to this standard set of weak reactions in \citet{Fischer:2016b} the (inverse)neutron decay is introduced, reaction (3) in Table~\ref{tab:nu-reactions}. This channel of opacity is enabled via the medium modifications with $Q=-Q_0+\triangle U$, which enters the expression of the elastic approximation. It results in the slight enhancement of the low-energy $\bar\nu_e$-opacity towards increasing density where $\triangle U>Q_0$, however, with a negligible impact on the overall supernova dynamics and neutrino signal.

\subsection{Nucleon-nucleon bremsstrahlung}
Neutrino-pair production (p) and absorption (a) are treated via their corresponding reaction kernels $\mathcal{R}_{\nu\bar\nu NN}^{\rm p/a}(\triangle E,\cos\theta)$. They relate initial and final states, depending on the sum of neutrino and antineutrino energies ($\triangle E = E_\nu+E_{\bar\nu}$) as well as on the momentum scattering angle $\cos\theta$ \cite[cf. Eq.~(5) in][]{Fischer:2012a}. It depends in turn on the lateral angles ($\mu_\nu=\cos\vartheta_\nu,\mu_{\bar\nu}=\cos\vartheta_{\bar\nu}$) and the relative azimuthal angle ($\phi$) between $\nu$ and $\bar\nu$. The corresponding collision term of the Boltzmann transport equation for neutrinos is given in \citet{Fischer:2012a} (assuming here $\hbar=c=1)$,
\begin{widetext}
\begin{eqnarray} \label{eq:boltz}
\left.\frac{df_\nu(E_\nu,\mu_\nu)}{dt}\right\vert_{\nu\bar\nu NN}
&=&
\left(1-f_\nu(E_\nu,\mu_\nu)\right) \int_0^\infty dE_{\bar\nu} E_{\bar\nu}^2 \int_{-1}^{+1} d\mu_{\bar\nu} \int_0^{2\pi} d\phi \,\, \mathcal{R}_{\nu\bar\nu NN}^\text{p}(-\triangle E,\cos\theta) \left(1-f_{\bar\nu}(E_{\bar\nu},\mu_{\bar\nu})\right)
\nonumber
\\
&-&
f_\nu(E_\nu,\mu_\nu) \int_0^\infty dE_{\bar\nu} E_{\bar\nu}^2 \int_{-1}^{+1} d\mu_{\bar\nu} \int_0^{2\pi} d\phi \,\,\mathcal{R}_{\nu\bar\nu NN}^\text{a}(\triangle E,\cos\theta) f_{\bar\nu}(E_{\bar\nu},\mu_{\bar\nu})~,
\end{eqnarray}
\end{widetext}

integrating the antineutrino phase space, it includes the appropriate (anti)neutrino occupation numbers $f_\nu(f_{\bar\nu})$ as well as final state blocking for the neutrino-pair production. Note further that the thermal and chemical equilibrium obtained under supernova conditions enables the relation between neutrino-pair emission and absorption kernels in Eq.~\eqref{eq:boltz} via detailed balance as follows,
\begin{eqnarray}
\mathcal{R}_{\nu\bar\nu NN}^\text{a}= \exp\left\{\triangle E/T\right\} \mathcal{R}_{\nu\bar\nu NN}^\text{p}.
\end{eqnarray}

At low densities the $NN$-interaction is governed by the OPE with the associated matrix element $\mathcal{M}$, for which in turn the relation to the neutrino-pair emission kernel is given in \citet{Hannestad:1997gc},
\begin{eqnarray}
\mathcal{R}_{\nu\bar\nu NN}^{\rm p}(-\triangle E,\cos\theta)\propto \langle\vert\mathcal{M}\vert^2\rangle \left(3-\cos\theta\right)~,
\label{eq:kernel0}
\end{eqnarray}
modulo normalization and multiplicative phase-space factors. The momentum integration of the nucleon occupation numbers, including final state Pauli blocking, is included in the matrix element for which the averaging $\langle\rangle$ is over all spins. Moreover, it has been shown in \citet{Friman:1979} that for non-relativistic nucleons, which is typical for temperatures below $T\sim 50$~MeV, only the axial-vector current contributes to the neutrino-pair processes from $NN$--bremsstrahlung,
\begin{eqnarray}
\langle\vert\mathcal{M}\vert^2\rangle \propto G_F^2 g_A^2~,
\label{eq:M}
\end{eqnarray}
with Fermi constant $G_F$ and axial-vector coupling constant $g_A$ with vacuum value, $g_A=1.27$. Analytical expressions for \eqref{eq:M} are provided in \citet{Hannestad:1997gc} to the astrophysics community based on the FOPE.

In order to account for the lowest-order medium dependence within the OPE approach, following \citet{Migdal:1990} two modifications can be considered: {\em (a)} dressing of the local interaction $\pi NN$-vertex ($\gamma$),
\setlength{\unitlength}{1.1mm}
\bigskip
\begin{eqnarray}
&&
\parbox{20mm}{\begin{fmffile}{vertex}
\begin{fmfchar*}(15,10)
\fmfleft{fb,f}
\fmfright{Z}
\fmfstraight
\fmf{dashes,label=$\pi$}{Z,v}
\fmf{fermion}{f,v,fb}
\fmffreeze
\fmflabel{$N$}{fb}
\fmflabel{$N$}{f}
\end{fmfchar*}
\end{fmffile}}
%
\;\;\;\;\;\scalebox{1.5}{$\longrightarrow$}\;\;\;\;\;\;\;\;\;\;
%
\parbox{20mm}{\begin{fmffile}{vertex_medium}
\begin{fmfchar*}(15,10)
\fmfleftn{l}{2}
\fmfright{r}
\fmfforce{(0.0w,0.0h)}{l1}
\fmfforce{(0.0w,1.0h)}{l2}
\fmfforce{(1.0w,0.5h)}{r}
\fmfpoly{full}{pr,pl2,pl1}
\fmfforce{(0.7w,0.5h)}{pr}
\fmfforce{(0.3w,0.7h)}{pl2}
\fmfforce{(0.3w,0.3h)}{pl1}
\fmf{fermion}{l2,pl2}
\fmf{fermion}{pl1,l1}
\fmf{dashes,label=$\pi$}{pr,r}
\fmflabel{$N$}{l1}
\fmflabel{$N$}{l2}
\end{fmfchar*}
\end{fmffile}}
,\\
&&
\nonumber
\label{eq:vertex}
\end{eqnarray}
and {\em (b)} modifications of the $\pi$-propagator ($D_\pi$). Based on the Fermi-liquid model, explicit expressions for {\em (a)} and {\em (b)} have been derived in \citet{Voskresenskaya:2001}. In the latter article it has also been shown that for the medium modifications of $D_\pi$ the most important contribution to the charged current processes (modified Urca) originates from neutrino reactions of intermediate nucleon-loops coupled to the $\pi$ for $\rho\gtrsim \rho_0/2$. However, these reaction channels are forbidden by symmetry reasons for $NN$--bremsstrahlung processes on neutral currents.

The medium modifications denoted via $(*)$ also alter the OPE matrix element, $\mathcal{M}\longrightarrow\mathcal{M}^{*}\propto G_F g_A^* \gamma D_\pi^* \gamma$. Neglecting furthermore contributions from the in-medium pion-propagator, i.e. assuming $D_\pi^{*}/D_{\pi} \simeq 1$, the ratio of MVOPE-to-FOPE matrix element is given as follows,
\begin{eqnarray}
\frac{\langle\vert\mathcal{M^*}\vert^2\rangle}{\langle\vert\mathcal{M}\vert^2\rangle} \simeq \left(\frac{g_A^*}{g_A}\right)^2\gamma^4~.
\label{eq:ratio}
\end{eqnarray}
Within the Fermi-liquid approach the $\pi NN$-vertex function $\gamma(\omega, k, p_F)$ -- depending on energy $\omega$, momentum $k$ and Fermi momentum $p_F$ -- is dressed by the inclusion of the $NN$ correlations according to expression~\eqref{eq:vertex} such that,
\begin{eqnarray}
\gamma(0,k,p_F)\simeq (1+C\,p_F)^{-1}~,
\end{eqnarray}
where the constant $C$ is expressed in terms of the appropriate Landau-Migdal parameter. Moreover, \citet{Voskresensky:1987} and \citet{Voskresenskaya:2001} provided the following numerical estimate,
\begin{eqnarray}
\gamma \simeq \left\{1 + \frac{1}{3}\left(\frac{m_N^*}{m_N}\right)\left(\frac{p_F(\rho)}{p_F(\rho_0)}\right)\right\}^{-1}~.
\label{eq:gamma}
\end{eqnarray}
%
Note that contributions from the effective nucleon mass ($m_N^*$) can be neglected in expression~\eqref{eq:gamma}, i.e. $m_N^*\simeq m_N$ for the densities relevant for neutrino transport and decoupling in core-collapse supernova simulations, $\rho< 10^{14}$~g~cm$^{-3}$. 

\begin{figure}[t!]
\centering
\includegraphics[width=0.975\columnwidth]{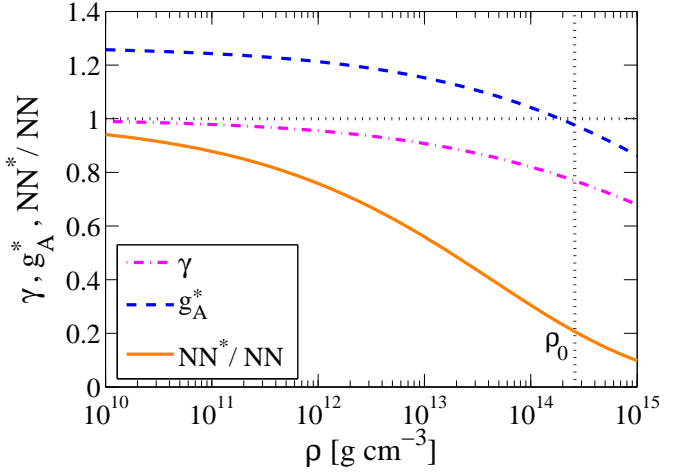}
\caption{(Color online) Medium dependence of the $\pi NN$-vertex, axial-vector coupling constant and the ratio Eq.~\eqref{eq:kernel}.}
\label{fig:gpiNN}
\end{figure}

Moreover,  the Goldberger-Treiman relation ensures the partial conservation of axial currents,
\begin{eqnarray}
m_N^{(*)} g_A^{(*)} = f_\pi^{(*)} g_{\pi NN}^{(*)}~,
\label{eq:GT}
\end{eqnarray}
with pion decay constant $f_\pi$, pion-nucleon coupling constant $g_{\pi NN}$ and nucleon mass $m_N$; it connects chiral physics and weak axial-vector currents. Relation~\eqref{eq:GT} holds in vacuum as well as in the nuclear medium, indicated via $(*)$. Applying the \citet{Brown:1991} scaling, $m_N^*/m_N \simeq f_\pi^*/f_\pi$, the ratio of the medium modified to the vacuum Goldberger-Treiman relation~\eqref{eq:GT} ensures the same scaling for axial-vector and $\pi NN$ coupling constants; it leads to the following dependence,
\begin{eqnarray}
g_A^*/g_A\simeq g_{\pi NN}^*/g_{\pi NN}\simeq\gamma~,
\label{eq:ga_lin}
\end{eqnarray}
which results in $g_A^*(\rho_0)\simeq 1$ as illustrated in Fig.~\ref{fig:gpiNN}. This medium dependent axial-vector coupling constant  is then employed in all other neutrino-matter interaction channels that involve axial-vector interactions.

Finally, according to the above expressions~\eqref{eq:kernel0}--\eqref{eq:ga_lin} the medium dependence of the neutrino-pair emission kernel scales as follows,
\begin{eqnarray}
\frac{\mathcal{R}_{\nu\bar\nu NN}^*}{\mathcal{R}_{\nu\bar\nu NN}} \simeq \left(\frac{1}{1+1/3\,(\rho/\rho_0)^{1/3}}\right)^6~,
\label{eq:kernel}
\end{eqnarray}
where the Fermi momenta are expressed in terms of the matter density, $p_F\propto\rho^{1/3}$.  In the following the medium modified process according to Eq.~\eqref{eq:kernel} is labelled as NN$^*$, while for the unmodified FOPE rate NN is used. Fig.~\ref{fig:gpiNN} illustrates the density dependence of $g_A^*$, being in good quantitative agreement with alternative approaches based on the low-density limit of chiral physics \citep[cf.][]{Friman:1999,Rho:2001,Meissner:2002,Carter:2002}, together with $\gamma$ and the ratio \eqref{eq:kernel} labelled as NN$^*$/NN. The overall onset of reduction of the process starts at densities of about $10^{12}$~g~cm$^{-3}$ with a suppression of about a factor of 5 at $\rho_0$, being in quantitative agreement with calculations of \citet{Bartl:2014} based on the $T$-matrix approach as well as chiral effective field theory.

\subsection{Neutrino opacity and mean-free path}
The neutrino opacity for pair processes, i.e. the ($\nu,\bar\nu$) absorption rate, is given in \citet{Fischer:2012a} obtained by integrating the reaction kernel taking into account the initial-state antineutrino phase space as follows,
\begin{eqnarray}
&& \chi_{\nu\bar\nu N N}(E_\nu,\mu_\nu) = \int_0^\infty dE_{\bar\nu} E_{\bar\nu}^2 \int_{-1}^{+1} d\mu_{\bar\nu} \int_0^{2\pi} d\phi
\nonumber \\
&& \;\;\;\;\;\;\;\;\;\;\;\;\;\;\;\;\;\;\;\
\times\;\mathcal{R}^{\rm a}_{\nu\bar\nu NN}(\triangle E,\cos\theta) \, f_{\bar\nu}(E_{\bar\nu},\mu_{\bar\nu})~.\;\;\;\;\;\;\;\;
\label{eq:chi}
\end{eqnarray}
For other weak processes, e.g., scattering on nucleons and electrons, a similar expression for the opacity $\chi$ as Eq.~\eqref{eq:chi} can be obtained via the integration of the corresponding scattering reaction kernels; they connect initial and final neutrino states similar as for $\nu\bar\nu NN$. For scattering the antineutrino occupation number $f_{\bar\nu}$ in expression~\eqref{eq:chi} has to be replaced with the final state neutrino occupation number. On the other hand, exceptions are the charged current absorption reactions which depend only on the incoming neutrino energy $E_\nu$ (within the elastic approximation employed here).

In order to study the medium dependence it is useful to define the inverse mean-free path as spectrum averaged quantity for each neutrino flavour ($\nu$) and weak process separately, based on the corresponding opacity channel $\chi_j(E_\nu,\mu_\nu)$. Therefore one has to integrate the remaining neutrino phase space as follows,
\begin{eqnarray}
\left\langle1/\lambda_j\right\rangle = \frac{2\pi}{n_\nu} \int dE_\nu E_\nu^2\,d\mu_\nu\,f_\nu(E_\nu,\mu_\nu)\,\chi_j(E_\nu,\mu_\nu)~,
\label{eq:mfp}
\end{eqnarray}
with the local neutrino number density $n_\nu$ and initial state neutrino occupation number $f_\nu$ corresponding to $n_\nu$. The integration of the neutrino phase space leaves only local dependencies of $\langle 1/\lambda \rangle$, e.g., in terms of the supernova state of matter given in terms of radial profiles.

\begin{figure}[t!]
\centering
\includegraphics[width=1.0\columnwidth]{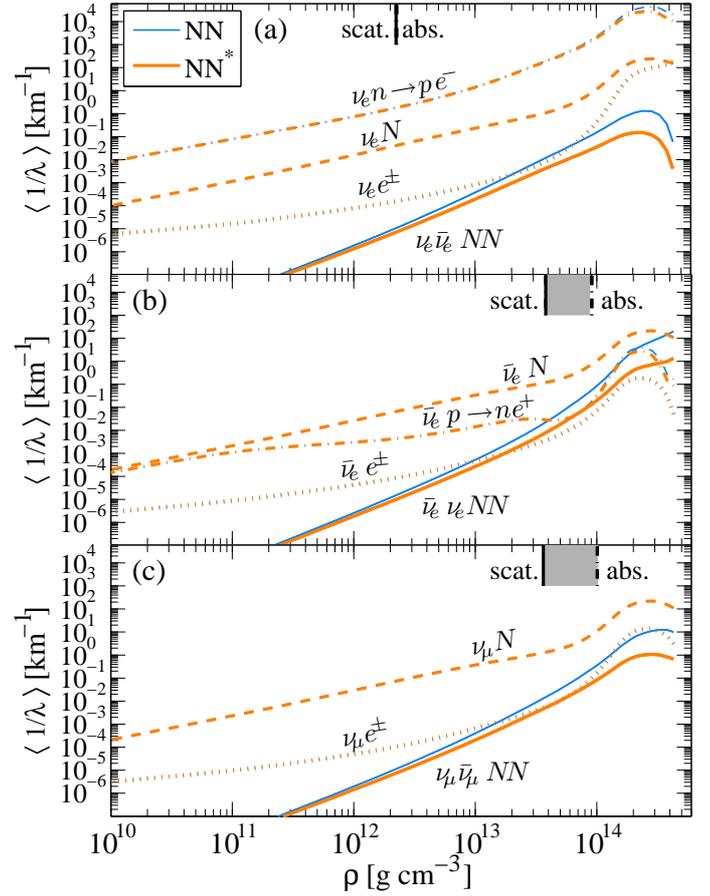}
\caption{(Color online) Density dependence of the average neutrino opacity~\eqref{eq:mfp} for selected weak processes, comparing $NN$ (thin blue lines) and $NN^*$ (thick brown lines); see text for definitions. The corresponding conditions are shown in Fig.~\ref{fig:dentye}.}
\label{fig:mfp}
\end{figure}
\begin{figure}[tb!]
\centering
\includegraphics[width=1.0\columnwidth]{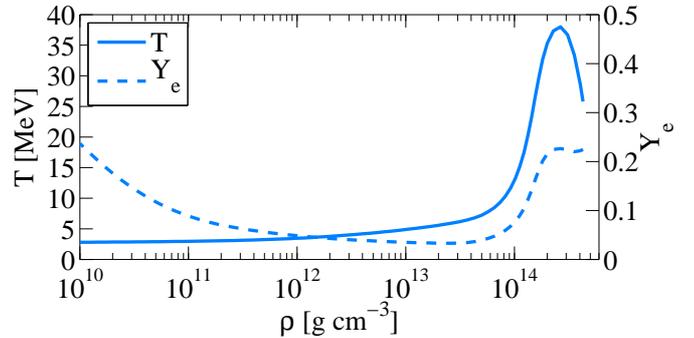}
\caption{(Color online) Profiles of temperature and $Y_e$ of the early PNS deleptonization phase at about 3~s post bounce.}
\label{fig:dentye}
\end{figure}

In Fig.~\ref{fig:mfp} the density dependence of the opacity \eqref{eq:mfp} is shown for several weak reaction channels $j$ with the largest contributions to the total opacity -- charged current absorptions (dash-dotted lines) -- elastic scattering on nucleons ($\nu N$, dashed lines) -- thermalization via inelastic scattering on electrons and positrons ($\nu e^\pm$, dotted lines) --  neutrino-pair absorption via $NN$--bremsstrahlung ($\nu\bar\nu N N$, solid lines). The associated conditions are shown in Fig.~\ref{fig:dentye} corresponding to the early PNS deleptonization phase at about 3~s post bounce. For $\nu_e$, illustrated in Fig.~\ref{fig:mfp}(a), the largest opacity is due to $\nu_e$-absorption on neutrons followed by elastic scattering on neutrons, since the neutrons as targets for these processes are the most abundant nuclear species. The opacity for all other weak processes is strongly suppressed. This also implies that $NN$--bremsstrahlung plays a negligible role for $\nu_e$ and that medium modifications NN$^*$ are unlikely to impact the overall $\nu_e$ spectra and fluxes. The situation is different for $\bar\nu_e$ as shown in Fig.~\ref{fig:mfp}(b), where the largest contribution to the opacity is due to elastic scattering on neutrons. The absorption of $\bar\nu_e$ on protons is heavily suppressed because protons are much less abundant than neutrons by about one order of magnitude and because medium modifications \citep[treated here at the mean level following][]{MartinezPinedo:2012} suppress the $\bar\nu_e$-absorption rate at low energies $E_{\bar\nu_e}<Q$. Note that the latter aspect is compensated by the inverse-neutron decay (reaction (3) in Table~\ref{tab:nu-reactions}) for which there is only a finite rate for $E_{\bar\nu_e}<Q$. Concerning the other inelastic channels, the opacity for $NN$--bremsstrahlung exceeds only slightly the opacity for $\bar\nu_e$-absorption on protons based on NN. With the medium dependent suppression NN$^*$ both opacities become on the same order in the region of neutrino decoupling illustrated via the grey band in Fig.~\ref{fig:mfp}(b) -- it will be further discussed below. This shifts back more relevance to the charged-current channel and is expected to modify the $\bar\nu_e$ properties. Note that for the $NN$--bremsstrahlung opacity as well as for inverse mean free path the neutrino spectra are used that are obtained from the supernova simulations that are based on Boltzmann neutrino transport. The situation for $\nu_{\mu/\tau}$ and $\bar\nu_{\mu/\tau}$, as illustrated in Fig.~\ref{fig:mfp}(c), is similar to the one for $\bar\nu_e$. Note that, unlike for $\bar\nu_e$, there is no charged-current absorption reaction for the heavy-lepton flavor neutrinos considered. Also here the largest contribution to the opacity comes from scattering on free neutrons. Inelastic contributions to the opacity are only due to $NN$--bremsstrahlung and scattering on electrons/positrons, which are strongly suppressed compared to scattering on neutrons. Now with the medium suppression NN$^*$ the opacity for $NN$--bremsstrahlung even drops below the one for inelastic scattering on electrons/positrons in the region of neutrino decoupling. Hence, the largest impact from the medium modifications of $NN$--bremsstrahlung is expected for the spectra of $\nu_{\mu/\tau}$ and $\bar\nu_{\mu/\tau}$.

Note also the small suppression of the other weak processes in Fig.~\ref{fig:mfp}, in particular for the charged-current absorption reactions. This is due to the employed medium dependence of $g_A^*$ according to expression~\eqref{eq:ga_lin}.

To guide the eye in Fig.~\ref{fig:mfp} short vertical black lines at the top of graphs~(a)--(c) mark the locations of the averaged neutrinospheres of last absorption labelled ''abs.'' (dash-dotted lines) and last elastic scattering labelled ''scat.'' (solid lines). For a definition of the neutrinospheres, see sec.~II.B in \citet{Fischer:2012a}. According to \citet{Raffelt:2001}, the region between ''scat.'' and ''abs.'' represents a scattering atmosphere (grey shaded region in Fig.~\ref{fig:mfp}) in which neutrinos transfer only momentum before finally reaching the free-streaming condition above ''scat.''. Note that unlike for $\bar\nu_e$ and $\nu_{\mu/\tau}$, for $\nu_e$ there is no scattering atmosphere because the opacity is dominated by charged-exchange absorption on neutrons by one order of magnitude above scattering on neutrons.

\section{PNS evolution}
\label{PNSevol}
In order to study the impact of medium modifications for the neutrino-pair processes from $NN$--bremsstrahlung in simulations of the PNS deleptonization, the rate according to expression~\eqref{eq:kernel} is implemented into the Boltzmann neutrino transport module of AGILE-BOLTZTRAN. The simulations are launched from the 18~M$_\odot$ pre-collapse progenitor of \citet{Woosley:2002zz}. In \cite{Fischer:2016} it was evolved consistently through all phases prior to the supernova explosion onset. The onset of the neutrino-driven supernova explosion -- shock revival during the post-bounce mass accretion phase -- is triggered following \citet{Fischer:2009af} via enhanced $\nu_e$ and $\bar\nu_e$ heating in the gain region, i.e. where net heating is established. It results in the explosion onset with the continuous shock expansion to increasingly larger radii at about 350~ms post bounce. This scenario for the shock revival is necessary since in general neutrino-driven explosions cannot be obtained in spherically symmetric simulations. Exceptions are low-mass stellar progenitors on the order of 8--10~M$_\odot$ \citep[cf.][]{Kitaura:2006,Melson:2015}. Moreover, it is not feasible to simulate the PNS deleptonization up to 10--30~s within multi-dimensional neutrino radiation hydrodynamics simulations. Note further that once the explosion proceeds the standard charged-current rates are switched back on.

\begin{figure}[t!]
\centering
\includegraphics[width=\columnwidth]{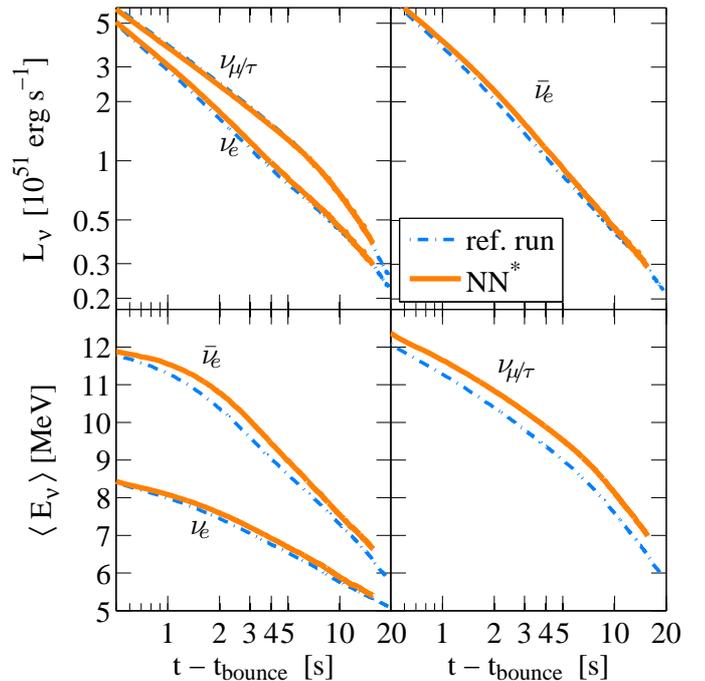}
\caption{(Color online) Evolution of neutrino luminosities and average energies.}
\label{fig:lumin}
\end{figure}

In the following, two different simulation setups will be discussed; {\bf ref.~run}: unmodified FOPE rate for $NN$--bremsstrahlung and vacuum $g_A$, compared to ${\bf NN^*}$: medium dependent suppression of $NN$--bremsstrahlung MVOPE and $g_A^*$ based on Eqs.~\eqref{eq:ga_lin} and \eqref{eq:kernel}. Both simulations use the EoS HS(DD2). The PNS deleptonization proceeds qualitatively similar in both simulations; the temperature decreases continuously at the PNS surface in the course of lepton losses and the entire PNS contracts accordingly with continuously increasing central density \citep[for illustration, cf. Figs.~(3) and (7) in][]{Fischer:2012a}. Note that results of the ref.~run have been published in \citet{Fischer:2016a}.

Quantitative differences arise in the neutrino signal illustrated in Fig.~\ref{fig:lumin}. For the simulation based on NN$^*$ the average energies of $\nu_{\mu/\tau}$ and $\bar\nu_e$ are increased up to about 1~MeV, compared to the reference case. For $\nu_e$ the effect is significantly weaker since the neutrino-pair processes act only as perturbation to the total opacity (see Fig.~\ref{fig:mfp}). The corresponding enhancement of the neutrino luminosities, in particular for $\nu_e$ and $\bar\nu_e$, is partly due to the feedback from $\nu_{\mu/\tau}$ via the neutrino-pair process (10) of Table~\ref{tab:nu-reactions}.

This situation remains during the entire PNS deleptonization up to 30~s post bounce, where towards later times the opacity from $\bar\nu_e$-absorption on protons reduces further due to the rising neutron excess and the increasing importance of the medium modifications for this weak process. At even late times, on the order of $t>40$~s post bounce, the transition is reached from neutrino opaque to transparent due to the continuously reducing temperatures below $T\simeq 1$~MeV, associated with neutrino decoupling from all densities including even above $\rho_0$; in order to simulate this transition other weak reaction, such as the (medium) modified Urca processes, drive the further evolution of the PNS towards the final cold neutron star. However, this extends beyond the scopes of the present study.

Note that the density dependent suppression for the neutrino-pair emission/production rate from $NN$--bremsstrahlung obtained here (see Fig.~\ref{fig:mfp}) is in quantitative agreement with \citet{Bartl:2014}, comparing $T$-matrix approach and chiral effective field theory with FOPE. However, the large enhancement of this rate found also in \citet{Bartl:2014} based on the $T$-matrix approach towards low density -- on the order of $\rho<10^{11}$~g~cm$^{-3}$ -- renders irrelevant because at such low density {\em (a)} the opacity is dominated by other weak processes for all neutrino flavors, as illustrated in Fig.~\ref{fig:mfp}, and {\em (b)} the neutrinos have already decoupled from matter.

\begin{figure}[t!]
\centering
\includegraphics[width=0.975\columnwidth]{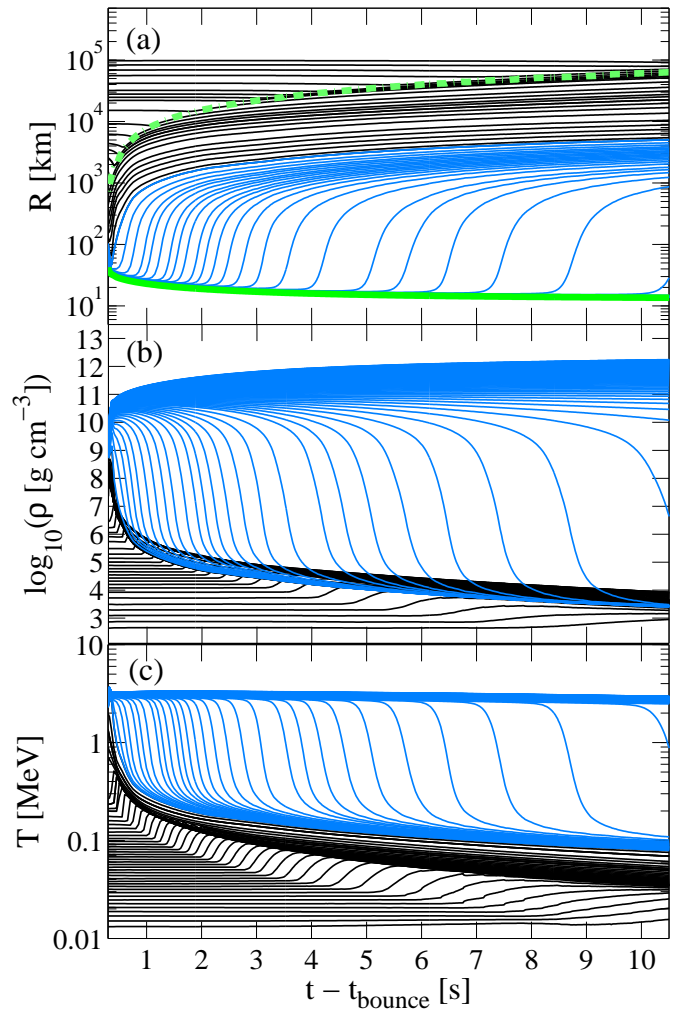}
\caption{(Color online) Evolution of selected tracer mass elements corresponding to dynamically ejected (black lines) material together with the shock expansion (green dash-dotted line) and the neutrino driven wind (blue lines) ejected from the PNS surface (green sold line).}
\label{fig:tracer}
\end{figure}

\section{Nucleosynthesis relevant conditions}
\label{ndw}
The increased spectral difference between $\bar\nu_e$ and $\nu_e$ found here for the simulation based on NN$^*$, compared to the ref.~run, has consequences for the nucleosynthesis of heavy elements associated with the low-mass outflow ejected via continuous neutrino heating from the PNS surface (solid green line in Fig.~\ref{fig:tracer}(a)) during the deleptonization -- known as neutrino driven wind. It has been studied based on the static wind equations \citep[cf.][]{Duncan:1986,Hoffman:1996aj,Thompson:2001ys} as well as within steady-state models \citep[cf.][]{Woosley:1992,Takahashi:1994yz} while \citet{Witti:1994} and \citet{Arcones:2006uq} were based on a hydrodynamics description with parametrized neutrino luminosities. In Fig.~\ref{fig:tracer} the post-bounce evolution of the neutrino driven wind (blue lines) is illustrated for selected tracer mass elements at the example of the reference case, as well as the material ejected dynamically (black lines) together with the expanding supernova shock illustrated via the green dash-dotted line in graph~(a). Differences of the dynamical evolution compared to the simulation based on NN$^*$ are negligible, continuously decreasing density and temperature as shown in Figs.~\ref{fig:tracer}(b) and \ref{fig:tracer}(c). The mass ejected in the neutrino driven wind is on the order of $6\times 10^{-4}$~M$_\odot$ during the first 10~s. Note that most material is ejected during the early evolution, $5.356\times 10^{-4}$~M$_\odot$ during the first 2~s. Hence, for the integrated nucleosynthesis analysis the early neutrino driven wind phase leaves the largest imprint on the yields.

In \cite{Qian:1996xt} it has been shown that the nucleosynthesis conditions of the neutrino driven wind are entirely determined from the neutrino fluxes and energies as well as their evolution. Moreover, in \citet{MartinezPinedo:2012} it has been found that the consistent treatment of charged current weak processes and nuclear EoS, which increases spectral differences between $\nu_e$ and $\bar\nu_e$, turns material to the neutron rich side with $Y_e\simeq0.47-0.49$ for the initial neutrino driven wind. However, with the inclusion of inelastic contributions and in particular weak magnetism corrections -- both are included here -- the neutron excess of the neutrino driven wind reduces significantly as shown in Fig.~\ref{fig:ndw}(a) for the reference run (blue dash-dotted line).

The inclusion of medium modifications at the level of NN$^*$, which increases slightly the difference between $\nu_e$ and $\bar\nu_e$ (see Fig.~\ref{fig:lumin}), turns back slightly the neutron excess during the early neutrino driven wind with minimum $Y_e\simeq 0.4895$ (see Fig.~\ref{fig:ndw}(a)). Material also stays slightly longer on the neutron rich side for about 1~s, before finally turning to proton rich conditions, which is due to the continuously reducing spectral differences between $\nu_e$ and $\bar\nu_e$ during the PNS deleptonization (see Fig.~\ref{fig:lumin}). The entropy per baryon, as additional nucleosynthesis relevant condition, is on the order of $s\simeq 50$~k$_{\rm B}$ per baryon during the early evolution, see Fig.~\ref{fig:ndw}~(b), rising only towards later times up to more than $s\sim 100$~k$_{\rm B}$ per baryon.

\begin{figure}[t!]
\centering
\includegraphics[width=0.975\columnwidth]{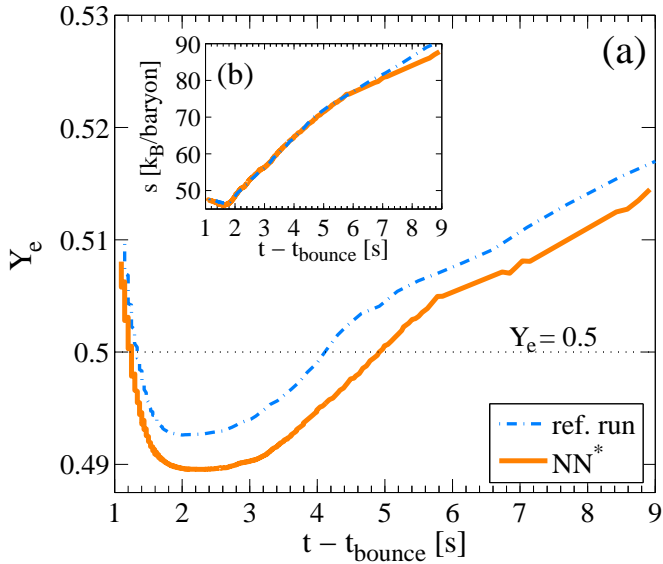}
\caption{(Color online) $Y_e$  and entropy sampled in the neutrino driven wind expansion at a radius of 1000~km, corresponding to the freeze-out conditions.}
\label{fig:ndw}
\end{figure}

Note that the entropies here are generally higher than reported previously due to the more massive 18~M$_\odot$ progenitor model explored here, compared to the lighter 11.2~M$_\odot$ progenitor which was explored in \citet{MartinezPinedo:2014}. The current PNS has a baryon mass of $M_{\rm B}=1.65$~M$_\odot$ compared to the much lighter PNS of $M_{\rm B}=1.19$~M$_\odot$ of \citet{MartinezPinedo:2014}. The relation between PNS mass and entropy of the neutrino driven wind is derived in \citet{Qian:1996xt}.

\section{Summary}
\label{summary}
Neutrino-pair emission and absorption from $NN$--bremsstrahlung is the relevant source and sink for heavy lepton flavor neutrinos above densities around $\rho_0/10$ in core-collapse supernova simulations. This article discusses the impact of medium modification of the vacuum OPE approximation for $NN$--bremsstrahlung, by means of providing a density-dependent parametrization of the medium modified $\pi NN$-vertex. The latter has been derived within the Fermi-liquid theory framework; details can be found in \citet{Migdal:1978}, \citet{Voskresensky:1987} and \citet{Migdal:1990}. It effectively suppresses the neutrino-pair emission rate with increasing density. The magnitude of suppression obtained here is in quantitative agreement with recent results of \citet{Bartl:2014} based on chiral effective field theory.

The modified neutrino-pair emission rate is then implemented into the Boltzmann neutrino transport module of the core-collapse supernova model AGILE-BOLTZTRAN in order to study the impact in simulations of the PNS deleptonization. Unlike $\nu_e$, which are dominated by charged-current absorption on neutrons, $\nu_{\mu/\tau}$ but also $\bar\nu_e$ are affected substantially. In particular, the reduced rate for neutrino-pair processes from $NN$--bremsstrahlung returns back more relevance to the charged-current $\bar\nu_e$-absorption on protons. This shifts the corresponding neutrinospheres for $\bar\nu_e$, and also for $\nu_{\mu/\tau}$, to somewhat higher density, which in turn results in enhanced neutrino fluxes and average energies.

These findings are in quantitative agreement with the results of \citet{Bartl:2016}, which are based on a parametrization of the study of \citet{Bartl:2014}, and have important consequences for the conditions relevant for the nucleosynthesis in the neutrino-driven wind, ejected from the PNS during the deleptonization. Note that inelastic and weak magnetism contributions reduce the neutron excess, from $Y_e\simeq 0.48$ of \citet{MartinezPinedo:2014} to $Y_e\simeq 0.495$, due to generally reduced spectral differences between $\nu_e$ and $\bar\nu_e$ for the latter. The simulation with the MVOPE approach increases back the spectral differences between $\nu_e$ and $\bar\nu_e$, which in turn increases the neutron excess slightly during the early neutrino driven wind phase. Material also stays longer on the neutron rich side. However, following \citet{Hoffman:1996aj}, the overall magnitude of $Y_e\simeq 0.49$ and the generally low entropy per baryon $s\simeq 50$~k$_{\rm B}$ indicate that the synthesis of heavy elements beyond Mo (with charge number $Z=52$) cannot be expected \citep[see also][]{Qian.Wasserburg:2007,Wanajo:2009}.

The here explored medium modifications for the $\pi NN$ vertex, and the consequent suppression of the neutrino-pair processes, are likely to affect also the early post-bounce mass accretion phase of core-collapse supernovae, i.e. prior to the possible onset of the supernova explosion which is studied to some extend in \citet{Bartl:2016}. In particular the losses associated with heavy-lepton flavor neutrinos may be reduced. Reduced cooling could then in turn attribute to more optimistic conditions for the revival of the stalled supernova shock. This remains to be explored in multi-dimensional core-collapse supernova simulations that are based on the sophisticated treatment of neutrino transport including a ''complete'' set of weak interactions.

Complementary may be the inclusion of inelastic neutrino nucleon-nucleon scattering, $\nu N N \leftrightarrows N N \nu$, in simulations of core-collapse supernovae and for the PNS deleptonization. Even though at finite temperatures on the order of 5--10~MeV the total opacity for all neutrino flavors will always be dominated by reactions with single neutrons as the most abundant nuclear species, this channel may substantially contribute to the energy-transfer for $\bar\nu_e$, $\nu_{\mu/\tau}$ and $\bar\nu_{\mu/\tau}$ in the region of neutrino decoupling, comparable to the one by neutrino-electron/positron scattering. Expressions for the $\nu N N \leftrightarrows N N \nu$ matrix element have been derived in \citet{Hannestad:1997gc} based on the FOPE approach, however, the correct inelastic dependence (momentum transfer) required for the scattering kernels of the collision integrals of the Boltzmann equation still has to be derived. This extends beyond the scopes of the present article and remains to be explored in future studies.

Note that this may become important towards later times of the PNS deleptonization when the core temperature drops below 1~MeV, entering the transition from neutrino diffuse to freely streaming. At that stage the direct Urca processes (1) -- (3) in Table~\ref{tab:nu-reactions} will become suppressed and other weak channels, e.g, modified Urca processes will dominate the further evolution.

\section*{Acknowledgement}
Special thanks belongs to D.~Voskresensky, P.~M.~Lo and G.~Mart{\'i}nez-Pinedo for the many discussions and for proofreading of the manuscript. I am also grateful for the private communications with D.~Blaschke, T.~Kl{\"a}hn, and C.~Sasaki. The author acknowledges support by the Polish National Science Center (NCN) under grant number UMO-2013/11/D/ST2/02645. The supernova simulations were performed at the Center for Scientific Computing (CSC) host at the University of Frankfurt (Germany).


\begin{thebibliography}{0}
\expandafter\ifx\csname natexlab\endcsname\relax\def\natexlab#1{#1}\fi

\end{thebibliography}


\begin{thebibliography}{83}
\expandafter\ifx\csname natexlab\endcsname\relax\def\natexlab#1{#1}\fi

\bibitem[{{Aouissat} {et~al.}(1995){Aouissat}, {Rapp}, {Chanfray}, {Schuck}, \&
  {Wambach}}]{Wambach:1995}
{Aouissat}, Z., {Rapp}, R., {Chanfray}, G., {Schuck}, P., \& {Wambach}, J.
1995, \npa, 581, 471

\bibitem[{Arcones {et~al.}(2007)Arcones, Janka, \& Scheck}]{Arcones:2006uq}
Arcones, A., Janka, H.-T., \& Scheck, L.
2007, \aap, 467, 1227

\bibitem[{{Bartl} {et~al.}(2014){Bartl}, {Pethick}, \& {Schwenk}}]{Bartl:2014}
{Bartl}, A., {Pethick}, C.~J., \& {Schwenk}, A.
2014, \prl, 113, 081101

\bibitem[{{Bartl} {et~al.}(2016){Bartl}, {Bollig}, {Janka} \& {Schwenk}}]{Bartl:2016}
{Bartl}, A., {Bollig}, R, {Janka}, H-T., \& {Schwenk}, A.
2016, (private communications), Arxiv e-prints, nucl-th/1608.05037

\bibitem[{{Blaschke} {et~al.}(2004){Blaschke}, {Grigorian}, \&
  {Voskresensky}}]{Blaschke:2004}
{Blaschke}, D., {Grigorian}, H., \& {Voskresensky}, D.~N.
2004, A\&A., 424, 979

\bibitem[{{Blaschke} {et~al.}(2013){Blaschke}, {Grigorian}, \&
  {Voskresensky}}]{Blaschke:2013}
{Blaschke}, D., {Grigorian}, H., \& {Voskresensky}, D.~N.
2013, \prc, 88, 065805

\bibitem[{{Blaschke} {et~al.}(2012){Blaschke}, {Grigorian}, {Voskresensky}, \&
  {Weber}}]{Blaschke:2012}
{Blaschke}, D., {Grigorian}, H., {Voskresensky}, D.~N., \& {Weber}, F.
2012, \prc, 85, 022802

\bibitem[{{Blaschke} {et~al.}(1995){Blaschke}, {R{\"o}pke}, {Schulz},
  {Sedrakian}, \& {Voskresensky}}]{Blaschke:1995}
{Blaschke}, D., {R{\"o}pke}, G., {Schulz}, H., {Sedrakian}, A.~D., \& {Voskresensky}, D.~N.
1995, \mnras, 273, 596

\bibitem[{{Brown} \& {Rho}(1991)}]{Brown:1991}
{Brown}, G.~E. \& {Rho}, M.
1991, \prl, 66, 2720

\bibitem[{Bruenn(1985)}]{Bruenn:1985en}
Bruenn, S.~W.
1985, \apjs, 58, 771

\bibitem[{Buras {et~al.}(2003)Buras, Janka, Keil, Raffelt, \&
  Rampp}]{Buras:2002wt}
Buras, R., Janka, H.-T., Keil, M.~T., Raffelt, G.~G., \& Rampp, M.
2003, \apj, 587, 320

\bibitem[{{Carter} \& {Prakash}(2002)}]{Carter:2002}
{Carter}, G.~W. \& {Prakash}, M.
2002, \plb, 525, 249

\bibitem[{{Duncan} {et~al.}(1986){Duncan}, {Shapiro}, \& {Wasserman}}]{Duncan:1986}
{Duncan}, R.~C., {Shapiro}, S.~L., \& {Wasserman}, I.
1986, \apj, 309, 141

\bibitem[{{Fischer} {et~al.}(2009){Fischer}, {Whitehouse}, {Mezzacappa},
  {Thielemann}, \& {Liebend{\"o}rfer}}]{Fischer:2009}
{Fischer}, T., {Whitehouse}, S.~C., {Mezzacappa}, A., {Thielemann}, F.-K., \& {Liebend{\"o}rfer}, M.
2009, \aap, 499, 1

\bibitem[{Fischer {et~al.}(2010)Fischer, Whitehouse, Mezzacappa, Thielemann, \& Liebend{\"o}rfer}]{Fischer:2009af}
Fischer, T., Whitehouse, S., Mezzacappa, A., Thielemann, F.-K., \& Liebend{\"o}rfer, M.
2010, \aap, 517, A80

\bibitem[{{Fischer} {et~al.}(2012){Fischer}, {Mart{\'{\i}}nez-Pinedo}, {Hempel}, \& {Liebend{\"o}rfer}}]{Fischer:2012a}
{Fischer}, T., {Mart{\'{\i}}nez-Pinedo}, G., {Hempel}, M., \& {Liebend{\"o}rfer}, M.
2012, \prd, 85, 083003

\bibitem[{{Fischer} {et~al.}(2013){Fischer}, {Langanke}, \& {Mart{\'{\i}}nez-Pinedo}}]{Fischer:2013}
{Fischer}, T., {Langanke}, K., \& {Mart{\'{\i}}nez-Pinedo}, G.
2013, \prc, 88, 065804

\bibitem[{{Fischer} {et~al.}(2014){Fischer}, {Hempel}, {Sagert}, {Suwa}, \& {Schaffner-Bielich}}]{Fischer:2014}
{Fischer}, T., {Hempel}, M., {Sagert}, I., {Suwa}, Y., \& {Schaffner-Bielich}, J.
2014, \epja, 50, 46

\bibitem[{{Fischer}(2016)}]{Fischer:2016}
{Fischer}, T.
2016, \epja, 52, 7

\bibitem[{{Fischer} {et~al.}(2016a){Fischer}, {Chakraborty}, {Giannotti}, {Mirizzi}, {Payez},  \&{Ringwald}}]{Fischer:2016a}
{Fischer}, T., {Chakraborty}, S., {Giannotti}, M., {Mirizzi}, A., {Payez}, A., \& {Ringwald}, A.
2016, \prd\,(submitted), ArXiv e-prints, astro-ph.HE/1605.08780

\bibitem[{{Fischer} {et~al.}(2016b){Fischer}, {Wu}, {Mart{\'{\i}}nez-Pinedo}, \& {Qian}}]{Fischer:2016b}
{Fischer}, T., {Wu}, M.-R., {Mart{\'{\i}}nez-Pinedo}, G., \& {Qian}, Y.-Z.
2016, (in preparation)

\bibitem[{{Friman} {et~al.}(1999){Friman}, {Rho}, \& {Song}}]{Friman:1999}
{Friman}, B., {Rho}, M., \& {Song}, C.
1999, \prc, 59, 3357

\bibitem[{{Friman} \& {Maxwell}(1979)}]{Friman:1979}
{Friman}, B.~L. \& {Maxwell}, O.~V. 1979,
\apj, 232, 541

\bibitem[{{Fuller} \& {Meyer}(1991)}]{Fuller:1991}
{Fuller}, G.~M. \& {Meyer}, B.~S. 1991,
\apj, 376, 701

\bibitem[{{Grigorian} \& {Voskresensky}(2005)}]{Grigorian:2005}
{Grigorian}, H. \& {Voskresensky}, D.~N. 2005,
\aap, 444, 913

\bibitem[{{Hanhart} {et~al.}(2001){Hanhart}, {Phillips}, \&
  {Reddy}}]{Reddy:2001}
{Hanhart}, C., {Phillips}, D.~R., \& {Reddy}, S.
2001, \plb, 499, 9

\bibitem[{Hannestad \& Raffelt(1998)}]{Hannestad:1997gc}
Hannestad, S. \& Raffelt, G.
1998, \apj, 507, 339

\bibitem[{{Hebeler} \& {Schwenk}(2010)}]{Hebeler:2010a}
{Hebeler}, K. \& {Schwenk}, A.
2010, \prc, 82, 014314

\bibitem[{Hempel \& Schaffner-Bielich(2010)}]{Hempel:2009mc}
Hempel, M. \& Schaffner-Bielich, J.
2010, \npa, 837, 210

\bibitem[{{Hempel}(2015)}]{Hempel:2015b}
{Hempel}, M.
2015 \prc, 91, 055807

\bibitem[{Hoffman {et~al.}(1997)Hoffman, Woosley, \& Qian}]{Hoffman:1996aj}
Hoffman, R., Woosley, S., \& Qian, Y.
1997, \apj, 482, 951

\bibitem[{Horowitz(2002)}]{Horowitz:2001xf}
Horowitz, C.
2002, \prd, 65, 043001

\bibitem[{{Horowitz} {et~al.}(2012){Horowitz}, {Shen}, {O'Connor}, \& {Ott}}]{Horowitz:2012}
{Horowitz}, C.~J., {Shen}, G., {O'Connor}, E., \& {Ott}, C.~D.
2012, \prc, 86, 065806

\bibitem[{H{\"u}depohl {et~al.}(2010)H{\"u}depohl, M{\"u}ller, Janka, Marek, \& Raffelt}]{Huedepohl:2010}
H{\"u}depohl, L., M{\"u}ller, B., Janka, H.-T., Marek, A., \& Raffelt, G.~G.
2010, \prl, 104, 251101

\bibitem[{{Janka} {et~al.}(1996){Janka}, {Keil}, {Raffelt}, \& {Seckel}}]{Janka:1996}
{Janka}, H.-T., {Keil}, W., {Raffelt}, G., \& {Seckel}, D.
1996, \prl, 76, 2621

\bibitem[{{Janka} {et~al.}(2007){Janka}, {Langanke}, {Marek}, {Mart{\'{\i}}nez-Pinedo}, \& {M{\"u}ller}}]{Janka:2007}
{Janka}, H.-T., {Langanke}, K., {Marek}, A., {Mart{\'{\i}}nez-Pinedo}, G., \& {M{\"u}ller}, B. 
2007, \physrep, 442, 38

\bibitem[{{Janka}(2012)}]{Janka:2012}
{Janka}, H.-T.
2012, Annual Review of Nuclear and Particle Science, 62, 407

\bibitem[{{Juodagalvis} {et~al.}(2010){Juodagalvis}, {Langanke}, {Hix}, {Mart{\'{\i}}nez-Pinedo}, \& {Sampaio}}]{Juodagalvis:2010}
{Juodagalvis}, A., {Langanke}, K., {Hix}, W.~R., {Mart{\'{\i}}nez-Pinedo}, G., \& {Sampaio}, J.~M.
2010, \npa, 848, 454

\bibitem[{Kitaura {et~al.}(2006)Kitaura, Janka, \& Hillebrandt}]{Kitaura:2006}
Kitaura, F., Janka, H.-T., \& Hillebrandt, W.
2006, \aap, 450, 345

\bibitem[{{Kr{\"u}ger} {et~al.}(2013){Kr{\"u}ger}, {Tews}, {Hebeler}, \& {Schwenk}}]{Krueger:2013}
{Kr{\"u}ger}, T., {Tews}, I., {Hebeler}, K., \& {Schwenk}, A.
2013, \prc, 88, 025802

\bibitem[{{Lattimer} \& {Lim}(2013)}]{Lattimer:2013}
{Lattimer}, J.~M. \& {Lim}, Y.
2013, \apj, 771, 51

\bibitem[{Liebend\"orfer {et~al.}(2001)Liebend\"orfer, Mezzacappa, \&
  Thielemann}]{Liebendoerfer:2001a}
Liebend\"orfer, M., Mezzacappa, A., \& Thielemann, F.-K.
2001, \prd, 63, 104003

\bibitem[{Liebend{\"o}rfer {et~al.}(2001)Liebend{\"o}rfer, Mezzacappa, Thielemann, Messer, Hix, {et~al.}}]{Liebendoerfer:2001b}
Liebend{\"o}rfer, M., Mezzacappa, A., Thielemann, F.-K., {et~al.}
2001, \prd, 63, 103004

\bibitem[{Liebendoerfer {et~al.}(2002)Liebendoerfer, Rosswog, \& Thielemann}]{Liebendoerfer:2002}
Liebend\"orfer, M., Rosswog, S., \& Thielemann, F.-K.
2002, \apjs, 141, 229

\bibitem[{Liebend\"orfer {et~al.}(2004)Liebend\"orfer, Messer, Mezzacappa, Bruenn, Cardall, {et~al.}}]{Liebendoerfer:2004}
Liebend\"orfer, M., Messer, O., Mezzacappa, A., {et~al.}
2004, \apjs, 150, 263

\bibitem[{Liebendoerfer {et~al.}(2005)Liebendoerfer, Rampp, Janka, \&  Mezzacappa}]{Liebendoerfer:2005a}
Liebend\"orfer, M., Rampp, M., Janka, H.-T., \& Mezzacappa, A.
2005, \apj, 620, 840

\bibitem[{{Mart{\'{\i}}nez-Pinedo} {et~al.}(2012){Mart{\'{\i}}nez-Pinedo},
  {Fischer}, {Lohs}, \& {Huther}}]{MartinezPinedo:2012}
{Mart{\'{\i}}nez-Pinedo}, G., {Fischer}, T., {Lohs}, A., \& {Huther}, L.
2012, \prl, 109, 251104

\bibitem[{{Mart{\'{\i}}nez-Pinedo} {et~al.}(2014){Mart{\'{\i}}nez-Pinedo}, {Fischer}, \& {Huther}}]{MartinezPinedo:2014}
{Mart{\'{\i}}nez-Pinedo}, G., {Fischer}, T., \& {Huther}, L.
2014, J.~Phys.~G, 41, 044008

\bibitem[{{Mei{\ss}ner} {et~al.}(2002){Mei{\ss}ner}, {Oller}, \&
  {Wirzba}}]{Meissner:2002}
{Mei{\ss}ner}, U.-G., {Oller}, J.~A., \& {Wirzba}, A.
2002, Annals of Physics, 297, 27

\bibitem[{{Melson} {et~al.}(2015){Melson}, {Janka}, \& {Marek}}]{Melson:2015}
{Melson}, T., {Janka}, H.-T., \& {Marek}, A.
2015, \apj, 801, L24

\bibitem[{Mezzacappa \& Bruenn(1993{\natexlab{a}})}]{Mezzacappa:1993gm}
Mezzacappa, A. \& Bruenn, S.~W.
1993{\natexlab{a}}, \apj, 405, 637

\bibitem[{Mezzacappa \& Bruenn(1993{\natexlab{b}})}]{Mezzacappa:1993gx}
Mezzacappa, A. \& Bruenn, S.~W.
1993{\natexlab{b}}, \apj, 410, 740

\bibitem[{{Migdal}(1978)}]{Migdal:1978}
{Migdal}, A.~B. 
1978, Rev.~Mod.~Phys., 50, 107

\bibitem[{{Migdal} {et~al.}(1990){Migdal}, {Saperstein}, {Troitsky}, \& {Voskresensky}}]{Migdal:1990}
{Migdal}, A.~B., {Saperstein}, E.~E., {Troitsky}, M.~A., \& {Voskresensky}, D.~N.
1990, \physrep, 192, 179

\bibitem[{Otsuki {et~al.}(1999)Otsuki, Tagoshi, Kajino, \& Wanajo}]{Otsuki:1999kb}
Otsuki, K., Tagoshi, H., Kajino, T., \& Wanajo, S.-y.
1999, \apj, 533, 424

\bibitem[{{Page} \& {Reddy}(2006)}]{Reddy:2006}
{Page}, D. \& {Reddy}, S.
2006, Annual Review of Nuclear and Particle Science, 56, 327

\bibitem[{Pons {et~al.}(1999)Pons, Reddy, Prakash, Lattimer, \& Miralles}]{Pons:1998mm}
Pons, J., Reddy, S., Prakash, M., Lattimer, J., \& Miralles, J.
1999, \apj, 513, 780

\bibitem[{Qian \& Woosley(1996)}]{Qian:1996xt}
Qian, Y.~Z. \& Woosley, S.
1996, \apj, 471, 331

\bibitem[{{Qian} \& {Wasserburg}(2007)}]{Qian.Wasserburg:2007}
{Qian}, Y.~Z. \& {Wasserburg}, G.~J.
2007, \physrep, 442, 237

\bibitem[{{Raffelt}(2001)}]{Raffelt:2001}
{Raffelt}, G.~G.
2001, \apj, 561, 890

\bibitem[{{Rapp} {et~al.}(1996){Rapp}, {Durso}, \& {Wambach}}]{Rapp:1996}
{Rapp}, R., {Durso}, J.~W., \& {Wambach}, J.
1996, \npa, 596, 436

\bibitem[{{Rapp} {et~al.}(1997){Rapp}, {Durso}, \& {Wambach}}]{Rapp:1997}
{Rapp}, R., {Durso}, J.~W., \& {Wambach}, J.
1997, \npa, 615, 501

\bibitem[{Reddy {et~al.}(1998)Reddy, Prakash, \& Lattimer}]{Reddy:1998}
Reddy, S., Prakash, M., \& Lattimer, J.~M.
1998, \prd, 58, 013009

\bibitem[{{Rho}(2001)}]{Rho:2001}
{Rho}, M. 2001,
Physics of Atomic Nuclei, 64, 637

\bibitem[{{Roberts} {et~al.}(2012){Roberts}, {Reddy}, \& {Shen}}]{Roberts:2012}
{Roberts}, L.~F., {Reddy}, S., \& {Shen}, G.
2012, \prc, 86, 065803

\bibitem[{{Schaab} {et~al.}(1997){Schaab}, {Voskresensky}, {Sedrakian}, {Weber}, \& {Weigel}}]{Voskresensky:1997}
{Schaab}, C., {Voskresensky}, D., {Sedrakian}, A.~D., {Weber}, F., \& {Weigel},  M.~K.
1997, \aap, 321, 591

\bibitem[{{Senatorov} \& {Voskresensky}(1987)}]{Voskresensky:1987}
{Senatorov}, A.~V. \& {Voskresensky}, D.~N.
1987, \plb, 184, 119

\bibitem[{Sigl(1996)}]{Sigl:1995ac}
Sigl, G.
1996, \prl 76, 2625

\bibitem[{Takahashi {et~al.}(1994)Takahashi, Witti, \& Janka}]{Takahashi:1994yz}
Takahashi, K., Witti, J., \& Janka, H.-Th.
1994, \aap, 286, 857

\bibitem[{{Tews} {et~al.}(2013){Tews}, {Kr{\"u}ger}, {Hebeler}, \& {Schwenk}}]{Tews:2013}
{Tews}, I., {Kr{\"u}ger}, T., {Hebeler}, K., \& {Schwenk}, A.
2013, \prl, 110, 032504

\bibitem[{Thompson {et~al.}(2001)Thompson, Burrows, \& Meyer}]{Thompson:2001ys}
Thompson, T.~A., Burrows, A., \& Meyer, B.~S.
2001, \apj, 562, 887

\bibitem[{Timmes \& Arnett(1999)}]{Timmes:1999}
Timmes, F.~X. \& Arnett, D.
1999, \apjs, 125, 277

\bibitem[{Typel {et~al.}(2013)Typel, Oertel, \& Klaehn}]{Typel:2013rza}
Typel, S., Oertel, M., \& Klaehn, T. 2013

\bibitem[{Typel {et~al.}(2010)Typel, Ropke, Klahn, Blaschke, \&
  Wolter}]{Typel:2009sy}
Typel, S., Ropke, G., Klahn, T., Blaschke, D., \& Wolter, H.
2010, \prc, 81, 015803

\bibitem[{{Voskresensky}(2001)}]{Voskresenskaya:2001}
{Voskresensky}, D.~N. 2001, in Lecture Notes in Physics, Berlin Springer
  Verlag, Vol. 578, Physics of Neutron Star Interiors, ed. D.~{Blaschke}, N.~K.
  {Glendenning}, \& A.~{Sedrakian}, 467

\bibitem[{Wanajo(2006{\natexlab{a}})}]{Wanajo:2006mq}
Wanajo, S.
2006{\natexlab{a}}, \apj, 650, L79

\bibitem[{Wanajo(2006{\natexlab{b}})}]{Wanajo:2006ec}
Wanajo, S.
2006{\natexlab{b}}, \apj, 647, 1323

\bibitem[{{Wanajo} {et~al.}(2009){Wanajo}, {Nomoto}, {Janka}, {Kitaura}, \& {M{\"u}ller}}]{Wanajo:2009}
{Wanajo}, S., {Nomoto}, K., {Janka}, H.-Th., {Kitaura}, F.~S., \& {M{\"u}ller}, B.
2009, \apj, 695, 208

\bibitem[{{Witti} {et~al.}(1994){Witti}, {Janka}, \& {Takahashi}}]{Witti:1994}
{Witti}, J., {Janka}, H.-Th., \& {Takahashi}, K.
1994, \aap, 286, 841

\bibitem[{Woosley {et~al.}(2002)Woosley, Heger, \& Weaver}]{Woosley:2002zz}
Woosley, S., Heger, A., \& Weaver, T.
2002, Rev.~Mod.~Phys., 74, 1015

\bibitem[{{Woosley} \& {Baron}(1992)}]{Woosley:1992}
{Woosley}, S.~E. \& {Baron}, E.
1992, \apj, 391, 228

\bibitem[{Woosley {et~al.}(1994)Woosley, Wilson, Mathews, Hoffman, \& Meyer}]{Woosley:1994ux}
Woosley, S., Wilson, J., Mathews, G., Hoffman, R., \& Meyer, B.
1994, \apj, 433, 229

\bibitem[{{Wu} {et~al.}(2015){Wu}, {Qian}, {Mart{\'{\i}}nez-Pinedo}, {Fischer}, \& {Huther}}]{Wu:2015}
{Wu}, M.-R., {Qian}, Y.-Z., {Mart{\'{\i}}nez-Pinedo}, G., {Fischer}, T., \& {Huther}, L.
2015, \prd, 91, 065016

\bibitem[{{Yakovlev} {et~al.}(2001){Yakovlev}, {Kaminker}, {Gnedin}, \&  {Haensel}}]{Yakovlev:2001}
{Yakovlev}, D.~G., {Kaminker}, A.~D., {Gnedin}, O.~Y., \& {Haensel}, P.
2001, \physrep, 354, 1

\bibitem[{{Yakovlev} {et~al.}(2005){Yakovlev}, {Gnedin}, {Gusakov}, {Kaminker}, {Levenfish}, \& {Potekhin}}]{Yakovlev:2005}
{Yakovlev}, D.~G., {Gnedin}, O.~Y., {Gusakov}, M.~E., {et~al.}
2005, \npa, 752, 590

\end{thebibliography}

\end{document}